\documentclass[proof]{WileyASNA-v1}

\articletype{Article Type}%
\usepackage{natbib}

\received{}
\revised{}
\accepted{}

\begin{document}

%\title{No inner companion to the hot Jupiter HATS-50b \protect\thanks{This is an example for title footnote.}}
%\title{No confirmation of the potential transiting companion of the hot Jupiter HATS-50b \protect\thanks{This is an example for title footnote.}}
\title{No further evidence for a transiting inner companion to the hot Jupiter HATS-50b \protect\thanks{Based on data obtained with the STELLA robotic telescopes in Tenerife, an AIP facility jointly operated by AIP and IAC.}}

\author{Matthias Mallonn}

\authormark{M. Mallonn}

\address[ ]{Leibniz-Institut f\"{u}r Astrophysik Potsdam (AIP), An der Sternwarte 16, 14482 Potsdam, Germany }

%\address[3]{\orgdiv{Org Division}, \orgname{Org Name}, \orgaddress{\state{State name}, \country{Country name}}}

\corres{\email{mmallonn@aip.de}}

%\presentaddress{This is sample for present address text this is sample for present address text}

\abstract{Most hot Jupiter exoplanets do not have a nearby planetary companion in their planetary system. One remarkable exception is the system of WASP-47 with an inner and outer nearby companion to a hot Jupiter, providing detailed constrains on its formation history. In this work, we follow-up on a tentative photometric signal of a transiting inner companion to the hot Jupiter HATS-50\,b. If confirmed, it would be the third case of a hot Jupiter with an inner companion. 63~hours of new ground-based photometry were employed to rule out this signal to about 96\,\% confidence. The injection of artificial transit signals showed the data to be of sufficient quality to reveal the potential photometric feature at high significance. However, no transit signal was found. The discrete pattern of observing blocks leaves a slight chance that the transit was missed. }


%\jnlcitation{\cname{%
%\author{Williams K.}, 
%\author{B. Hoskins}, 
%\author{R. Lee}, 
%\author{G. Masato}, and 
%\author{T. Woollings}} (\cyear{2016}), 
%\ctitle{A regime analysis of Atlantic winter jet variability applied to evaluate HadGEM3-GC2}, %\cjournal{Q.J.R. Meteorol. Soc.}, \cvol{2017;00:1--6}.}

%%\fundingInfo{Funding info text.}

\maketitle

%\footnotetext{\textbf{Abbreviations:} ANA, anti-nuclear antibodies; APC, antigen-presenting cells; IRF, interferon regulatory factor}

\section{Introduction}
\label{sec_intro}
Close-in Jupiter-sized exoplanets, the so-called hot Jupiters, are a rare phenomenon. Only about 1 percent of solar-like stars host such a planet of at least six~Earth radii in size and an orbital period below 10~days \citep{Mayor2011,Fressin2013,Wang2015}. On the contrary, it is the type of exoplanets which are best characterized \citep{Seager2010}. This is caused by the circumstance that the hot Jupiters produce the largest signal in radial velocity and photometric measurements. Thus, most of the observational effort to characterize the atmospheres of exoplanets by emission, transmission and phase-resolved spectroscopy concentrated on these targets due to the observational feasibility \citep[see reviews by][]{Crossfield2015,Deming2017,Sing2018}.

However, there are numerous unsolved questions concerning the hot Jupiters. One of these is the quest how they achieved their orbits. A review was recently provided by \cite{Dawson2018}. The scenarios discussed in the literature are that they either formed in situ \citep{Bodenheimer2000,Batygin2016,Lee2016} or that they migrated inward toward their current position by interactions with the protoplanetary discs \citep{Lin1996,Ida2008} or by high-eccentricity tidal migration \citep{Ida2013,Wu2011}. 

For the majority of hot Jupiters, the high-eccentricity tidal migration presents a feasible option \citep{Dawson2018}. It is broadly in agreement to the observed eccentricity distribution of hot Jupiters, and to the finding that hot Jupiters are generally alone in their close environment \citep{Steffen2012,Huang2016}. Companions orbiting the same host star have mostly been found only very far from the hot Jupiter in the outer parts of the planetary system \citep{Knutson2014,Ngo2015}, a few were found with a companion within 1~AU \citep{Butler1999,Wright2009,Endl2014,Hartman2014}. The discovery of multiple, closely packed planets in the system of WASP-47 presented a striking contrast \citep{Becker2015,Neveu2016}. Next to the hot Jupiter WASP-47\,b, the host star harbors two sub-Neptunes, one interior and one exterior to the hot Jupiter. The orbits of these near-by siblings would have become unstable during the high-eccentricity phase of the hot Jupiter, thus it likely formed through a different evolutionary path \citep{Weiss2017}. The only other planetary system known with an inner companion to a hot Jupiter is Kepler-730. The sub-Neptune Kepler-730\,c orbits the host star with a period of 2.8~days, while the hot Jupiter Kepler-730\,b has a period of 6.5~days \citep{Canas2019}. The discovery of similar planet systems would be very valuable to learn more about the occurrence rate of such exceptional systems and their evolution.

Recently, \cite{Henning2018} announced the discovery of four hot Jupiters. In the HAT-South photometry dataset of one of them, the system of HATS-50, they found an additional signal of a transiting planet at low signal-to-noise ratio. While the hot Jupiter HATS-50\,b has an orbital period of $\sim$\,3.38~days, the additional candidate signal showed a period of 0.766~days, a transit depth of 3.2~mmag, and a duration of 46~minutes. A transit fit by \cite{Henning2018} resulted in an object size larger than Neptune, with the short signal length suggesting a grazing transit. Because HATS-50\,b is not grazing, but fully transiting the host star, this putative configuration of the planetary system would exhibit a mutual inclination of more than 10~degrees. The radial velocity (RV) data, obtained to confirm the planetary nature of HATS-50\,b, showed substantial jitter, which might be related to the potential inner companion. No significant RV variation could be measured at the period of the photometric signal, thus \cite{Henning2018} derived an upper mass limit for the candidate HATS-50\,c of 0.16~$M_\mathrm{J}$. The space satellite TESS \citep{Ricker2015} would be well suited to confirm or reject the tentative transit signal. However, according to the Web TESS Viewing Tool\footnotetext{https://heasarc.gsfc.nasa.gov/cgi-bin/tess/webtess/wtv.py}, HATS-50 will not be observed by TESS throughout its first two years of operation.

In this work, we present our photometric follow-up campaign to verify the existence of the tentative shallow transit signal suggested by \cite{Henning2018}. While the ground-based detection of the targeted small signal is clearly challenging, meter-sized ground-based telescope have proven to be capable of detecting transit and eclipse signals of milli-magnitude amplitude \citep[e.g.,][]{Lendl2013,Lendl2017,West2019,Mallonn2019}.
In Section~2, we present the new ground-based photometric observations and their data reduction. Section~3 provides the analysis of a transit signal of the hot Jupiter HATS-50\,b. The analysis of the entire data set to search for the potential transiting inner companion to the hot Jupiter is given in Section~4. We conclude our work in Section~5.

\section{Observations and data reduction}
We obtained 63~hours of time-series photometry with the 1.2m robotic STELLA telescope and its wide field imager WiFSIP \citep{Strassmeier2004}. The instrument provides a field of view (FoV) of 22\,$'\times$ 22\,$'$ on a scale of 0.32$''$/pixel \citep{Granzer2010}. The detector is a single 4096$\times$4096 back-illuminated thinned CCD with 15$\mu$m pixels. The telescope is located on the northern hemisphere at the Canary Islands. HATS-50 is a southern target with a declination of $-26^{\circ}$. Seen from STELLA, the target is visible in the summer season for 3.5~hours at an airmass smaller than 2.0. %From previous time-series photometry, we gained the experience that high-precision photometry is possible up to this airmass limit. 
In total, we observed blocks of 3.5~hours in 18~nights. An observing log is given in Table~\ref{tab_overview}. Due to the uncertainty of the ephemeris of the suggested additional transit signal, we scheduled the observation only according to the availability of telescope time. Thus, the observations have been taken randomly in respect to the orbital phase of the putative planet.

In comparison to the population of known hot Jupiter host stars, HATS-50 is rather faint with $\mathrm{V} = 14$~mag. We compensated for this by an exposure time of 180~seconds for all observations, yielding an observing cadence of 223~seconds and only $\sim$\,12~data points per potential transit event. However, in average, each point in orbital phase is covered $\sim$\,3 times under the assumption of an orbital period of 0.766~days. All observations were taken in the Sloan r' filter, and we slightly defocussed the telescope to achieve more stable photometry. The average point-to-point scatter of the differential photometric light curves is 2.3~mmag.

\begin{table}
%\tiny
\caption{Overview of observations of HATS-50 taken with the STELLA telescope in the Sloan r' filter. The columns provide the observing date, the number of the observed individual data points, the exposure time, the dispersion of the data points as root-mean-square (rms), and the airmass range of the observations.}
\label{tab_overview}
\begin{center}
%\begin{tabular}{lccccccc}
\begin{tabular}{lccr}
\hline
\hline
\noalign{\smallskip}
Date & $N_{\mathrm{data}}$ &  rms (mmag) &  Airmass \\
\hline
\noalign{\smallskip}
2018-06-09  & 56  &  1.7   &  1.71 - 2.10   \\
2018-06-10  & 56  &  2.3   &  1.71 - 2.08   \\
2018-06-13  & 57  &  2.2   &  1.71 - 2.02   \\
2018-06-14  & 53  &  2.4   &  1.71 - 1.98   \\
2018-06-16  & 56  &  2.0   &  1.71 - 2.02   \\
2018-06-17  & 56  &  1.9   &  1.71 - 2.05   \\
2018-08-29  & 57  &  2.4   &  1.71 - 2.14   \\
2018-08-30  & 57  &  2.5   &  1.71 - 2.10   \\
2018-09-01  & 57  &  2.8   &  1.71 - 2.15   \\
2018-09-03  & 57  &  2.8   &  1.71 - 2.10   \\    
2018-09-04  & 56  &  2.3   &  1.71 - 2.08   \\
2018-09-06  & 57  &  2.3   &  1.71 - 2.03   \\
2018-09-07  & 56  &  2.4   &  1.71 - 2.02   \\
2018-09-08  & 57  &  2.0   &  1.71 - 2.00   \\ 
2018-09-09  & 57  &  2.9   &  1.71 - 1.99   \\
2018-09-10  & 57  &  2.2   &  1.71 - 1.99   \\ 
2018-09-11  & 57  &  2.0   &  1.71 - 2.01   \\
2018-09-16  & 57  &  2.1   &  1.71 - 2.10   \\

\hline                                                                                                     
\end{tabular}
\end{center}
\end{table}

%By chance, during the first observing block of the target, a partial transit of HATS-50b was caught.

The data reduction was done similarly to our previous analyses of exoplanet transit observations with STELLA/WiFSIP \citep[e.g.,][]{Mallonn2015,Mallonn2016}. In short, the bias and flat field correction was supplied by the official STELLA pipeline. For aperture photometry, we employed SExtractor \citep{Bertin1996}. For each individual observing night, we tested and applied the aperture size that minimized the scatter in the light curve. Also, the selection of multiple comparison stars was chosen nightly to minimize the scatter. As the last step of the extraction of the differential light curves, we applied a 4\,$\sigma$ clipping to the nightly data to remove outliers.

\section{Timing of the transit of HATS-50\,b}
\label{sec_b_transit}
\begin{figure}
 \centering
 \includegraphics[width=\hsize]{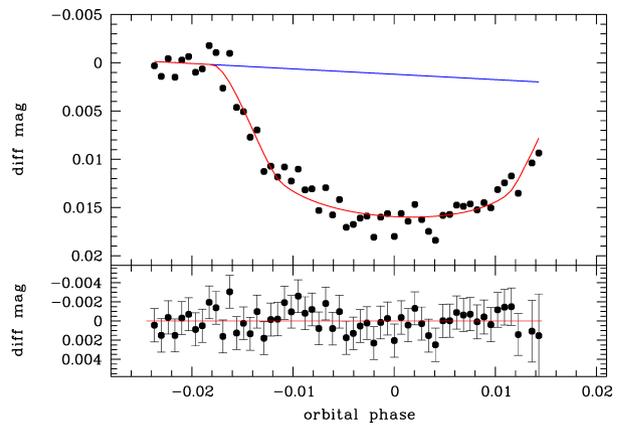}
 \caption{Transit light curve of the hot Jupiter HATS-50\,b. Upper panel: the red solid line shows the transit plus linear trend model, the blue solid lines shows only the linear trend. Black data points denote the STELLA observations. Lower panel: black data points show the light curve residuals after subtraction of the transit plus linear trend model.}
 \label{plot_fig1}
    % /work2/mamul/STELLA/HATS50/lit_lcs/makeplot_STELLA01.prg
\end{figure}

During the first observing block, a partial transit of HATS-50\,b was observed by chance. Its light curve is shown in Figure~\ref{plot_fig1}. We derive the mid time of the transit and provide it here for any future parameter and ephemeris refinement work on this hot Jupiter. 

%refine the orbital ephemeris of the hot Jupiter exoplanet. For this purpose, we include the transit follow-up photometry of \cite{Henning2018} in a re-analysis together with our new dataset. The literature data involve observations of six transit events, two complete transits and four partial ones. 
The transit analysis is done with the publicly available software tool JKTEBOP \citep{Southworth2004,Southworth2008}. The transit model involves the fit parameters of the orbital semi-major axis scaled by the stellar radius $a/R_{\star}$, the orbital inclination $i$, the planet-star radius ratio $k$, the mid point of the transit $T_0$, the orbital period $P$, the eccentricity of the orbit $e$, the argument of periastron $\omega$, and detrending coefficients $c_{0,1}$. For detrending, we chose a linear function over time which was found to minimize the Bayesian Information Criterion \citep[BIC,][]{Schwarz78}. This choice is in agreement to \cite{Mackebrandt2017}, who also found a first order polynomial as best choice for the detrending of photometric WiFSIP data of about three hours duration. We re-scaled the photometric uncertainties to match the point-to-point dispersion in the light curve (see Section~\ref{sec_anal}).
%Since the retrieved litertature light curves were already treated with decorrelation and detrending algorithms \citep{Henning2018}, we include a second-order polynomial over time in their re-analysis to include the detrending uncertainty in our error estimation.
The limb darkening coefficients of the quadratic limb darkening law for the Sloan~r' band were adopted from \cite{Henning2018}.

Since we provide in this work only one new partial transit observation of HATS-50\,b, we do not intend to refine transit parameters like $a/R_{\star}$ or $i$. %However, since our observation extends the observed time range by three more years, we attempt a refinement of the orbital period estimation. 
Thus, we fix all fit parameters to the values derived in \cite{Henning2018}, and fit the light curve for the transit timing $T_0$ and the two detrending coefficients. 

The resulting transit mid time of the hot Jupiter HATS-50\,b is $2458279.67541\pm0.0018$ (BJD$_{\mathrm{TDB}}$), which is in 1.3\,$\sigma$ agreement to the ephemeris of \cite{Henning2018}, if the uncertainties of the timing measurement and the ephemeris are quadratically combined. We subtracted the transit model of HATS-50\,b for further analysis from the light curve of the first observing block. No other data set of our sample was affected by a transit of the hot Jupiter.

%NOCHMAL checken, macht diese Zeit ueberhaupt Sinn, wurde da beobachtet?? 
%The seven individual timings are summarized in Table XXX. In the next step, we fit the seven data sets simultaneously for a value of $T_0$ and the orbital period $P$, still including a simultaneous individual detrending of each data set involved. The newly derived ephemeris is
%\begin{equation}
%T_n\,=\,T_0 + n \cdot P \, ,
%\end{equation}
%with $T_0$ as the timing zero point, $P$ as the orbital period, and $n$ as the number of epochs passed since $T_0$. XX All seven individual timings agree with this linear ephemeris to within 2~sigma XX.

%Henning:   56870.34870 0.00068   3.8297015   0.0000046  BJD TDB
%this work: 57360.55002 0.00154   3.8296914   0.000010
%-> period hat 1.0 sig diskrepanz
%Abweichung meines Datenpunktes von Henning ephemeris ist ca 1.4 sigma bei quadr combination of uncertainty transit and epehem...

\section{Analysis and results}
\label{sec_anal}
Before we search for a transit signal of an inner companion to HATS-50\,b in the data, we tested different parametric detrending functions and used the BIC to select the best. We compared linear combinations of low-order polynomials with time, airmass, object FWHM, and x and y position on the chip as independent variables. As in Section~\ref{sec_b_transit} for the first observing block, we found the very simple first-order polynomial over time to minimize the BIC also for all other observing blocks.

In the next step, we re-scale the individual photometric uncertainties. The values derived from SExtractor tend to underestimate the errors, since they only account for the photon noise of target and comparison star and their respective background. We inflate the individual photometric uncertainties by a common factor such that the light curves per observing block reach a reduced $\chi^2$ of unity versus the detrending function. 
%From this stage of the data analysis, we work on the detrended light curves.

\subsection{Initial transit fit}
To search for a transit signal of an inner companion to HATS-50\,b, first we perform a transit fit with JKTEBOP. In one attempt, we fit for all relevant parameters $a/R_{\star}$, $i$, $k$, $P$, $T_0$, and two detrending coefficients per observing block. The two limb darkening coefficients of the quadratic limb darkening law were fixed to the values of \cite{Henning2018}, and for simplicity we fixed the eccentricity of the potential planet c to zero. In another attempt, we first created a transit model that produced a grazing transit with $10^{\circ}$ mutual inclination to HATS-50\,b and a transit depth and duration as suggested by \cite{Henning2018}, and then fixed $a/R_{\star}$ and $i$ to the values of this model, leaving all other parameters free. In both fit versions, we used the values of \cite{Henning2018} for $P$ and $T_0$ as initial parameter values. Then, in additional fit runs, we moved the initial $T_0$ through the orbital phase in steps of 0.2.
In no case we detected a transit feature similar to the tentative transit detection of \cite{Henning2018}. The best fit models had transit depths of 1~mmag or smaller which were always in agreement to a depth of zero within 2\,$\sigma$ or less. The BIC values of the models with a transit were larger than the BIC values of a model only including detrending, thus the no-transit models were favored. Also, a visual inspection provided no indication for a transit feature of about 3~mmag depth.

%Free-to-fit parameters are XXX. The two limb darkening coefficients of the quadratic limb darkening law were fixed to the values of \cite{Henning2018}, and for simplicity we fixed the eccentricity of the potential planet c to zero. Initial parameter values are ... The best fit transit model has a period of bla and a transit depth of bla, which is consistent with a null detection. 

We can already conclude at this stage of our work that we cannot confirm the suggested transit signal with our new STELLA photometry. The remaining part of our work will be devoted to the question whether we can rule out this signal significantly.

\subsection{Recovery of an injected transit signal}
We will investigate whether the new STELLA photometry is of sufficient quality to reveal a transit signal as suggested by \cite{Henning2018}. For this purpose, we create a transit model of 46~minutes duration and a depth of 3.2~mmag with JKTEBOP and inject it into our observing data. We apply the suggested orbital period of 0.7662482~days and the timing zero point of $T_c\,=\,2455274.38586$. Afterwards, we move the injected transit signal through orbital phase in steps of 0.2, hence we repeat the exercise of signal recovery five times. The radius ratio $k$ of the injected transit was 0.06, and in all five cases, we recovered this value well with values ranging from 0.057 to 0.066 and an uncertainty range from 0.004 to 0.006. In all five cases, the BIC values favored the model including the transit.

\subsection{A simple box model for the transit search}
In the previous sections of this work, we have demonstrated that our data are of sufficient quality to reveal a transit feature of suggested depth, duration, and periodicity. However, the ability to reveal a photometric transit signal obviously depends on the orbital phase coverage. This phase coverage might not be complete for certain orbital periods. To investigate on the question whether our data rule out the 3~mmag transit feature for a range of orbital periods around the suggested value, we compare the BIC values of the transit-plus-detrending model versus the detrending-only model for a large number of combinations of $P$ and $T_0$. We consider a transit model to be rejected if its BIC value is larger by a difference of 10 than the corresponding detrending-only model. To save computational time, we approximate the transit signal with a simple box model. Because of the potentially grazing shape of the transit \citep{Henning2018}, we design a box of 46~minutes length, but a more shallow depth of 2.5~mmag instead of the suggested depth of 3.2~mmag. We chose a certain value for the orbital period and move the box through the orbital phase in steps of 0.01 (which corresponds to 11~minutes for a period of 0.766~days).

We consider a period range of $\pm \, 5\,\%$ of the suggested period as wide and exhaustive for our exercise, since the typical period uncertainty of a transiting, ground-based detected planet is $\ll \, 0.01\,\%$, however typically with substantially more follow-up observations. Thus, we examine a period range from 0.729 to 0.804~days in steps of 0.0001~days.

\begin{figure*}
 \centering
 \includegraphics[width=\hsize]{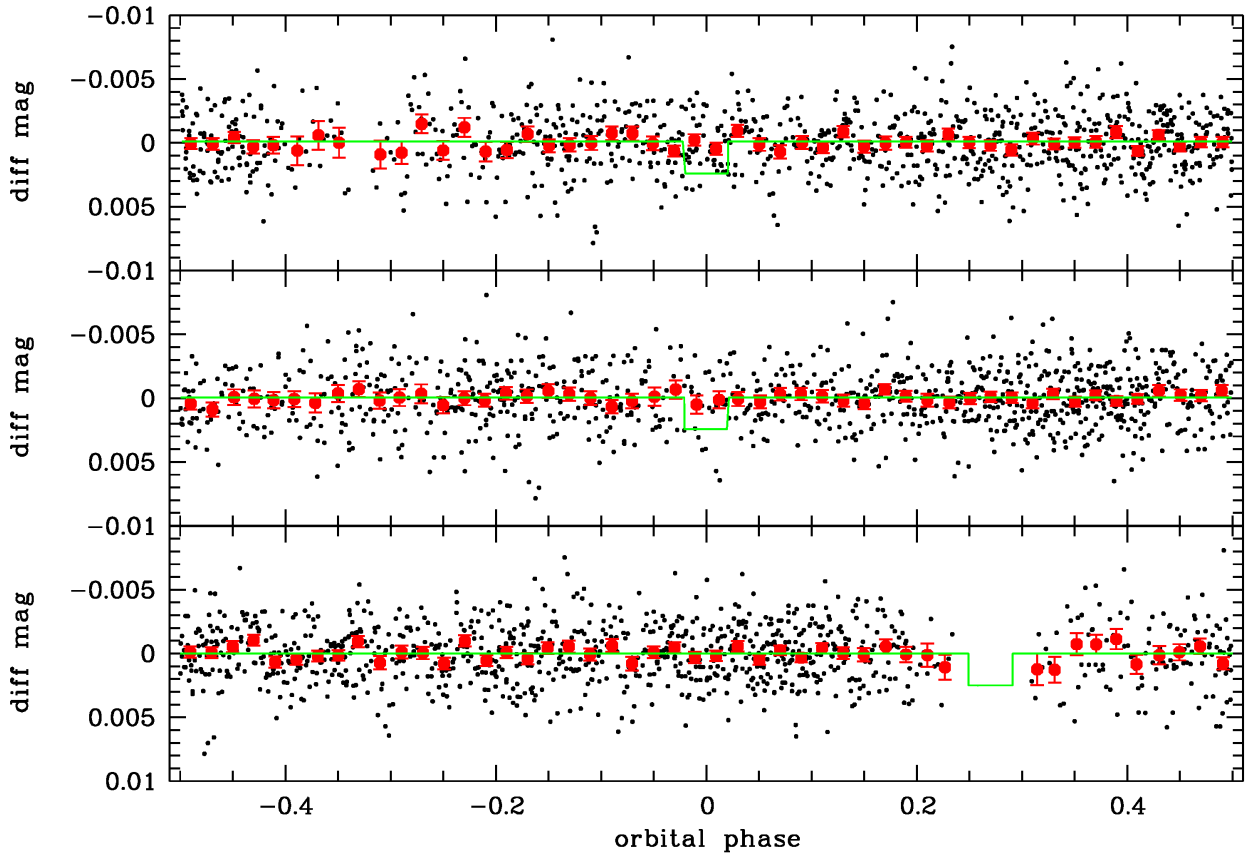}
 \caption{Phase-folded STELLA photometry. Upper panel: Shown are the observing data in black folded to the suggested period of \mbox{$P=0.766248$~days}. In red, the data are binned in steps of 0.05 in phase. The box model of 46~minutes length and 2.5~mmag depth is shown in green, centered on the transit mid time suggested by \cite{Henning2018}. Middle panel: The data are phase-folded to a period with complete phase coverage, here $P=0.7670$~days. The transit feature can be ruled out for all orbital phases. Lower panel: An example of a period with incomplete phase coverage, here $P=0.7656$~days, is given. The transit feature is ruled out for all orbital phases covered with observations, however, the feature might potentially be hidden in the phase gap, as illustrated by the green model.}
 \label{plot_fig2}
    % /work2/mamul/STELLA/HATS50/workon/plot_binned_lc_three.prg
\end{figure*}

The result is a distribution of BIC differences for each period-phase combination. The first outcome was that in no case, such difference between transit-plus-detrending and detrending-only model was below -10, i.e. in no case the transit model was significantly favored. On the other side, the transit feature was not ruled out in 100\,\% of the period-phase combinations, which means there are cases with a BIC difference of below 10 and no model was significantly favored over the other. We inspected examples of these near-zero BIC differences and found them all to be associated to an incomplete orbital phase coverage caused by our irregular sequence of observing blocks. In Figure~\ref{plot_fig2}, we show in the upper panel the STELLA data phase folded to the period of 0.766428 suggested by \cite{Henning2018}. In the middle panel, we show a randomly drawn example of a period for which we have full phase coverage. For such period, the transit feature is ruled out at all phases. In the lower panel of Figure~\ref{plot_fig2}, we give an example of an orbital period close to the suggested value, for which our data cannot rule out the transit feature completely because the phase coverage is not complete. Potentially, we could have missed the transit with our observations and it could be hidden in the phase gap.  

For the examined combinations of orbital period and phase from $P=0.729$ to $P=0.804$, we obtain a BIC value larger than 10 in 95.9\,\% of all cases. That means that there is a slight probability of $\sim$\,4\,\% that we have missed the transit. Hence, we can rule the transit feature out to about 96\,\% confidence. The value remains very similar for a period range tighter around the suggested period and varies from 95.5\,\% to 96.5\,\% dependent on the specific period interval.
%For a certain range in orbital period, we move this box model through the phase, and compare the BIC of the detrending plus box model to the pure detrending model. We consider the box model of a period-phase pair as rejected if the BIC is greater than 10 compared to the no-transit, pure detrending model.

%The flat line satisfies the non-variable flux of the host star because \cite{Henning2018} limited the host star variability to below $\sim$\,1~mmag by the HATSouth discovery photometry. 
%The box model is ruled out if its BIC value is larger by 10 than the BIC value of the flat line.

% 0.76624822days
%  depth of 3.2mmag, and duration of 46minutes
  
% beta factor \citep{Gillon06,Winn08}

\section{Discussion and conclusion}
We presented 63~hours of follow-up photometry of HATS-50 to verify a transiting feature of an inner companion to the hot Jupiter HATS-50\,b, which was tentatively detected by \cite{Henning2018}. The existence of an inner or nearby outer companion to a hot Jupiter is informative of its formation and evolution, since the three theoretically most discussed formation scenarios make distinct predictions. In situ formation can form nearby planets outside of orbital resonances, disk migration scenarios result in nearby resonant companions, and high-eccentricity migration eliminates nearby companions \citep{Dawson2018}. Observational evidence have shown that the vast majority of hot Jupiters does not have nearby companions \citep{Latham2011,Steffen2012,Huang2016}. Constraints on the evolutionary history come from statistical properties of their population rather than from individual planet systems. On the contrary, the only planetary systems found with an inner companion to a hot Jupiter so far, WASP-47 and Kepler-730, already allowed conclusions on their evolution based on the individual planetary parameters. For example, \cite{Weiss2017} described that none of the three formation scenarios mentioned above, in situ formation, disk migration, and high-eccentricity migration, can produce all the physical properties of the WASP-47 system. Instead, it is more likely that the system underwent multiple stages of planet formation with individual planets formed at different times. 

The discovery of other nearby companions to hot Jupiters or a tight constraint on their existence would therefore be of great interest for our understanding of hot Jupiter formation. Our new photometric data set of HATS-50 could not confirm the existence of the suggested transiting inner companion to HATS-50\,b, though the injection and recovery of artificial transit signals proved the data to be of sufficient photometric quality. We ruled out the existence of the suggested transit feature to 96\,\% confidence, leaving the slight possibility that we missed the transit due to our discrete observing pattern of many observing blocks. This pattern causes an incomplete phase coverage for a small fraction of the examined orbital period interval. Space-based observatories like Spitzer, TESS, or CHEOPS could deliver the continuous observation of a full planet orbit without phase gaps typical for ground-based observatories to rule out an inner transiting companion at even higher confidence. %However, we demonstrate in this work the value of ground-based observations with small-sized telescopes for exoplanet follow-up.    

%\backmatter

\section*{Acknowledgments}
We thank Carolina von Essen and Katja Poppenhaeger for helpful discussions on the manuscript. We thank Thomas Granzer for technical support with the STELLA observations. This work made use of the SIMBAD data base and VizieR catalog access tool, operated at CDS, Strasbourg, France, and of the NASA Astrophysics Data System (ADS).

%This work was performed under the auspices of the \fundingAgency{National Nuclear Security Administration} of the US Department of Energy at Los Alamos National Laboratory under Contract No. \fundingNumber{DE-AC52-06NA25396} and supported. The authors acknowledge the partial support of the \fundingAgency{DOE Office of Science ASCR Program} under Contract No. \fundingNumber{SJS-AC52-06NA25968}. This work was partially supported by the \fundingAgency{Czech Technical University} grant \fundingNumber{SGS16/247/OHK4/3T/14}, the \fundingAgency{Czech Science Foundation} project \fundingNumber{14-21318S} and by the \fundingAgency{Czech Ministry of Education} project \fundingNumber{RVO 68407700}.

%\subsection*{Author contributions}
%This is an author contribution text. This is an author contribution text. This is an author contribution text.  

%\subsection*{Financial disclosure}
%None reported.

%\subsection*{Conflict of interest}
%The authors declare no potential conflict of interests.

%\section*{Supporting information}
%The following supporting information is available as part of the online article:
%
%\noindent
%\textbf{Figure S1.}
%{500{\uns}hPa geopotential anomalies for GC2C calculated against the ERA Interim reanalysis. The period is 1989--2008.}
%
%\noindent
%\textbf{Figure S2.}
%{The SST anomalies for GC2C calculated against the observations (OIsst).}

%\appendix

%\nocite{*}% Show all bib entries - both cited and uncited; comment this line to view only cited bib entries;
\bibliography{mm_only_bib.bib}%

%\section*{Author Biography}
%(if applicable)

%\begin{biography}{\includegraphics[width=60pt,height=70pt,draft]{empty}}{\textbf{Author Name.} This is sample author biography text this is sample author biography text this is sample author biography text this is sample author biography text this is sample author biography text this is sample author biography text this is sample author biography text this is sample author biography text this is sample author biography text .}
%\end{biography}

\end{document}